\documentclass[12pt]{article}

\usepackage[numbers,sort&compress]{natbib}
\usepackage{graphicx}
\usepackage{amsmath}
\usepackage{amssymb}
\usepackage{hyperref}
\usepackage{feynmp}

\usepackage{ifpdf}
\ifpdf
\fi

\ifx\hypersetup\sadfkjashdfkxja\else
\hypersetup{pdftitle={Form Factors of Chiral Primary Operators at Two Loops
in ABJ(M)}}
\hypersetup{pdfauthor={Donovan Young}}
\hypersetup{pdfsubject={}}
\hypersetup{pdfkeywords={}}
\hypersetup{plainpages=false}
\hypersetup{bookmarksnumbered=true}
\hypersetup{pdfstartview=FitH}
\hypersetup{pdfpagemode=UseNone}
\hypersetup{colorlinks=false}
\hypersetup{citebordercolor={.5 1 .5}}
\hypersetup{urlbordercolor={.5 1 1}}
\hypersetup{linkbordercolor={1 .7 .7}}
\fi

\def\ni{\noindent}
\def\be{\begin{equation}}
\def\ee{\end{equation}}
\def\bsp{\be\begin{split}}
\def\la{\langle}
\def\ra{\rangle}
\def\dag{\dagger}

\def\lr{\leftrightarrow}

\def\G{\Gamma}

\def\a{\alpha}
\def\b{\beta}
\def\g{\gamma}

\def\d{\delta}
\def\e{\epsilon}
\def\m{\mu}

\def\Tr{\mbox{Tr}}
\def\n{\nu}

\def\r{\rho}
\def\l{\lambda}

\def\o{\omega}

\def\p{\partial}

\def\bZ {\mathbb{Z}}
\def\w{\wedge}

\makeatletter

\newcommand{\Rmnum}[1]{\expandafter\@slowromancap\romannumeral #1@}
\makeatother

\renewcommand{\title}[1]{\vbox{\center\LARGE{#1}}\vspace{5mm}}
\renewcommand{\author}[1]{\vbox{\center\large{#1}}\vspace{5mm}}
\newcommand{\address}[1]{\vbox{\center\em#1}}
\newcommand{\email}[1]{\vbox{\center\tt#1}\vspace{5mm}}

\begin{document}

\numberwithin{equation}{section} 
\bibliographystyle{utphys}
\newpage
\setcounter{page}{1}
\pagenumbering{arabic}
\renewcommand{\thefootnote}{\arabic{footnote}}
\setcounter{footnote}{0}

\begin{fmffile}{ff1}

\fmfcmd{%
style_def wiggarr expr p =
draw (wiggly p);
fill (arrow p)
enddef;}

\begin{titlepage}
\hfill {\tt NORDITA-2013-34}\\
\title{\vspace{1.0in} {\bf Form Factors of Chiral Primary Operators at Two Loops
in ABJ(M)}}
 
\author{Donovan Young}

\address{NORDITA\\ KTH Royal Institute of Technology and Stockholm
  University\\ Roslagstullsbacken 23, SE-10691 Stockholm, Sweden}

\email{donovany@nordita.org}

\abstract{We calculate the colour-ordered form factor for chiral
  primary operators built from $J$ scalar fields of ABJ(M) theory to
  $J$ scalar final states. We work in the 't Hooft limit and show that
  the leading quantum correction is ${\cal O}(\l^2)$, where $\l$ is
  the 't Hooft coupling. We evaluate this leading correction using
  standard Feynman diagrams and dimensional regularization, and find
  that the leading divergence is $1/\e^2$ where the spacetime
  dimension is $d=3-2\e$. We further find that the result respects
  maximal transcendentality.}

\end{titlepage}

\tableofcontents
\section{Introduction}

The study of scattering amplitudes has emerged as a new direction of
research in the AdS/CFT correspondence. The discoveries of the twistor
string description, BDS ansatz, null-polygonal Wilson loop
equivalence, recursion relations, unitarity-based techniques,
colour-kinematics duality, Yangian symmetry, Grassmannian formulation,
and the relation between gauge theory and gravity amplitudes have
together founded a new research area, sometimes referred to as
amplitudeology (see \cite{Beisert:2010jr} chapter \Rmnum{5} for a
review and partial list of references).

Traditionally, the main objects of interest on the CFT side of the
AdS/CFT correspondence have been gauge invariant local
operators. Non-local operators have also been studied. For example
Wilson loops have received wide attention, and surface operators have
also been investigated. But until relatively recently, scattering
amplitudes had been studied less. Indeed, it is unnatural to consider
scattering amplitudes in a CFT, since the absence of asymptotic states
precludes their definition. This can be overcome through dimensional
regularization: the introduction of an IR regularization $d=d_0-2\e$
with $\e<0$ allows amplitudes to be calculated and their divergences
as $\e\to 0$, at least in the case of ${\cal N}=4$ supersymmetric
Yang-Mills theory (SYM) in four dimensions, display a beautiful
structure captured by exponentiation and the appearance of the cusp
anomalous dimension \cite{Bern:2005iz}. On the string side of the
correspondence Alday and Maldacena \cite{Alday:2007hr} showed that
this IR regularization corresponds to the introduction of a brane deep
in the bulk of AdS on which the scattering is taking place. A
T-duality operation which inverts the bulk direction (and hence swaps
IR and UV divergences) maps the process to a fundamental string with
minimal embedding having a null-polygon on the AdS boundary as its own
boundary, i.e. the string-dual to a null-polygonal Wilson loop in the
gauge theory. The full extent of this T-duality symmetry is reflected
in dual-superconformal symmetry \cite{Drummond:2008vq}, where the
$PSU(2,2|4)$ symmetry group of ${\cal N}=4$ SYM is doubled, and a
Yangian symmetry emerges from this structure
\cite{Drummond:2009fd}. Since Yangians are fundamental structures in
integrable systems, it would appear that the famed integrability of
${\cal N}=4$ SYM as applied to the two-point functions of local
operators (i.e. the spectral problem) may be emerging also in the
study of scattering amplitudes.

It is highly desirable to make links with the program of integrability
outside the application to the spectral problem. In this spirit the
question of how integrability impacts higher-point correlation
functions of local operators, such as three-point functions, has begun
to receive attention. In the case of the $AdS_4/CFT_3$ correspondence
between ${\cal N}=6$ superconformal Chern-Simons theory (ABJM) and
M-theory on $AdS_4\times S^7/\bZ_k$ \cite{Aharony:2008ug}, the
question of three-point functions offers a unique opportunity. Both
the usual spectral-problem integrability and the Yangian invariance of
scattering amplitudes appears to be present in ABJM, and thus we can
also ask what impact integrability may have on the three-point
functions. The most basic local operators are the chiral primaries
(CPO's) -- these are symmetrized traces of scalar fields, and have
protected conformal dimensions. In the case of ${\cal N}=4$ the
three-point functions of the chiral primaries are also protected,
indeed they are independent of the coupling. The same, however, is not
true in ABJM. The three-point functions of chiral primaries have
non-trivial coupling dependence and an interesting structure of
contractions between the symmetric-traceless tensors which define
them\footnote{We also note that extremal n-point functions of CPO's
  have been computed in the free field theory limit of ABJ(M) in
  \cite{Caputa:2012dg}.}; at strong coupling supergravity indicates
that they scale as $\l^{1/4}/N$, where $\l=N/k$ is the ABJM 't Hooft
coupling \cite{Hirano:2012vz}. The interpolating function governing
the three-point function of CPO's in ABJM stands a good chance of
being a simple, fundamental quantity in the $AdS_4/CFT_3$
correspondence, and it would not be surprising to find that it is
related to something which integrability could compute. Of course, it
might also be related to the scaling function $h(\l)$, which is
undetermined by integrability, but might be provided via localization
techniques \cite{Correa:2012hh,Cardinali:2012ru}.

Form factors offer a bridge between traditional local gauge-invariant
operator correlation functions and scattering amplitudes, and could
help elucidate the connection between the integrable and other
structures found in each case. Perhaps more importantly, they are
easier to compute then correlation functions, and, as recently
demonstrated \cite{Engelund:2012re}, may be used to construct
higher-point correlation functions using generalized unitarity. The
form factor calculations presented here serve as a basis both to study
these objects further in their own right and as a tool to build
correlation functions, especially the three-point functions of CPO's,
in perturbation theory, which the author hopes to report upon in the
near future.

Form factors have been studied in the context of ${\cal N}=4$ SYM in a
series of papers. The founding paper \cite{vanNeerven:1985ja} studied
length-2 CPO form factors with two scalar final states to two loops,
where exponentiation of IR divergences, controlled by the cusp and
colinear anomalous dimensions as in the case of amplitudes, and
further exponentiation of the finite part was noted. Links to
Parke-Taylor structure at tree-level were developed in
\cite{Selivanov:1998hn}. More recently the length-2 CPO form factor
\cite{Brandhuber:2010ad,Brandhuber:2011tv,Brandhuber:2012vm} and
length-$n$ CPO form factor \cite{Bork:2010wf,Bork:2011cj} have been
studied with a general number of gluons in the final states in MHV and
more general configurations, where exponentiation of divergences
continues to be present, while a remainder function for the finite
part appears for the first time for three external states. This
remainder function has been evaluated at the two-loop level and
displays some intriguing connections to the maximally transcendental
part of an analogous QCD calculation \cite{Brandhuber:2012vm}.
Supersymmetrization of both the length-2 operator, in the sense of replacing it
with the entire stress-energy multiplet of operators, and the final
states has also been accomplished
\cite{Bork:2011cj,Brandhuber:2011tv}, while recursion relations and
(dual) MHV rules for the form factors have been developed in
\cite{Brandhuber:2011tv}. The form factor for the Konishi state has
also been considered in \cite{Bork:2010wf}.

In this paper we calculate the leading quantum correction in the 't
Hooft limit to the form factors for chiral primary operators of length
$J$, with $J$ scalar final states, in the ${\cal N}=6$ superconformal
Chern-Simons theory known as ABJ(M)
\cite{Aharony:2008ug,Aharony:2008gk}. We show that the leading
correction is at the two-loop level, and reduce the relevant Feynman
diagrams to master integrals. Our results are contained in
(\ref{rJ2ABJM}), (\ref{rJ2ABJ}) and (\ref{rJJABJM}), (\ref{rJJABJ})
for the $J=2$ and $J>2$ cases respectively. The $J=2$ case in the ABJM
theory has been concurrently computed via generalized unitarity in
\cite{Brandhuber:2013gda}. Our (\ref{rJ2ABJM}) matches this result to all orders in
$\e$. The form factors diverge as ${\cal O}(\e^{-2})$, and respect the
principle of maximal transcendentality.

The paper is organized as follows. In section \ref{sec:ff} we
introduce and define the form factors, show that the one-loop
correction vanishes, and calculate the two-loop result first
for the $J=2$ case, which has some special subtleties owing to the
colour structure of the diagrams, and then for the $J>2$ cases. We
conclude with a short discussion in section
\ref{sec:disc}. Conventions, Feynman rules, and some other details of
the calculation may be found in the appendix.
 
\section{ABJ(M) form factors}
\label{sec:ff}

We would like to compute the leading perturbative correction to the
form factor of a chiral primary operator built of scalar fields $Y^A$
and $Y^\dag_A$, given by
\be\label{defop}
{\cal O}^J_{A} = ({\cal C}_{A})^{A_1\ldots A_{J/2}}_{B_1\ldots B_{J/2}} \,
\Tr \left(Y^{B_1} Y^\dag_{A_1} \cdots Y^{B_{J/2}} Y^\dag_{A_{J/2}}\right),
\ee
where ${\cal C}_{A}$ is completely symmetric in upper and
(independently) in lower indices, while the trace taken on any pair
consisting of one upper and one lower index vanishes. The tensors are
orthonormal, so that
\be
({\cal C}_{A})^{I_1\ldots I_{J/2}}_{K_1\ldots K_{J/2}} 
({\cal C}^*_{B})^{K_1\ldots K_{J/2}}_{I_1\ldots I_{J/2}} =\delta_{AB} ,
\ee
and the two-point function, i.e. the conformal dimension of the
operator, is protected against quantum corrections by supersymmetry.

The colour-ordered form factor $F(\{s_{ij}\})$ we consider is for the production of
scalar final states, and is defined by the following
expression\footnote{We suppress the gauge group indices. To restore
  them note that $Y^{A_a}\to(Y^{A_a})_{i_a\hat i_a}$,
  $Y^\dag_{B_a}\to(Y^\dag_{B_a})_{\hat j_a j_a}$ and so the LHS of
  (\ref{ffdef}) should carry a factor of $\prod_{a=1}^{J/2} \d_{i_{a+1} j_a}
  \d_{\hat i_a \hat j_a}$, where $a\sim a+J/2$.}
\bsp\label{ffdef}
\frac{J}{2}\,({\cal C}_{A})^{A_1\ldots A_{J/2}}_{B_1\ldots B_{J/2}}
\,&F(\{s_{ij}\})\\
&\equiv
\Bigl\la Y^{A_1}(p_1) Y^\dag_{B_1}(p_2) \cdots Y^{A_{J/2}}(p_{J-1}) 
Y^\dag_{B_{J/2}}(p_J) \,\Bigl|\, {\cal O}_A^J(0)\, \Bigr|\, 0\,\Bigr\ra,
\end{split}
\ee
where $s_{ij}=(p_i+p_j)^2$ are the Mandelstam invariants associated
with the on-shell (i.e. $p_i^2=0$) external legs. This definition has
been chosen so that the tree-level result is $F(\{s_{ij}\})=1$. To
begin with we specialize to the $J=2$ case, as many of the diagrams
will simply be recycled across the extra legs in the $J>2$ cases.

\subsection{One loop is zero}

We represent the tree-level result in the following diagrammatic
language (e.g. for $J=2$)
\vspace{-2cm}
\begin{center}\[
\parbox{20mm}{\vspace{2cm}
\begin{fmfgraph*}(17,40)
\fmftop{v1}
\fmfleft{v2}
\fmfright{v3}
\fmfv{decoration.shape=circle,decoration.filled=30}{v1}
\fmf{plain_arrow}{v1,v2}
\fmf{plain_arrow}{v1,v3}
\fmflabel{$p_1$}{v2}
\fmflabel{$p_2$}{v3}
\end{fmfgraph*}}=1,\]
\end{center}
\vspace{-1.25cm}
so that the gray blob at the top represents the operator, and
final state momenta are outgoing. The one-loop correction to the form
factor vanishes. It is given by the one-gluon exchange between
adjacent legs, depicted below for the $J=2$ case
\vspace{-2cm}
\begin{center}\[
\parbox{20mm}{\vspace{2cm}
\begin{fmfgraph*}(17,40)
\fmftop{v1}
\fmfleft{v2}
\fmfright{v3}
\fmfv{decoration.shape=circle,decoration.filled=30}{v1}
\fmf{plain}{v1,vc1}
\fmf{plain}{vc1,v2}
\fmf{plain}{v1,vc2}
\fmf{plain}{vc2,v3}
\fmffreeze
\fmf{photon,right=0.15}{vc1,vc2}
\fmffreeze
\fmfposition
\end{fmfgraph*}}\propto \int \frac{d^3q}{(2\pi)^3} \frac{\e_{\m\n\r}
  \,q^\m p_1^\n p_2^\r}{q^2(q+p_1)^2(q-p_2)^2} = 0.\]
\end{center}
\vspace{-1.25cm}
This vanishes under the integration of the loop momentum because of
the integrand being an odd function of the loop momentum.  A more general
argument for the vanishing of these types of diagrams on the physical
dimension is given in appendix \ref{sec:appv}.

\subsection{Two-loop analysis: $J=2$ case}

We find that the following diagrams contribute at two-loop order
(wiggly line $=$ gluon, dashed line $=$ fermion, plain line $=$ scalar)
\vspace{1cm}
\begin{center}
\begin{tabular}[t]{cccc}
\parbox{20mm}{
\begin{fmfgraph*}(17,40)
\fmftop{v1}
\fmfleft{v2}
\fmfright{v3}
\fmf{phantom}{v1,vb}
\fmfv{decoration.shape=circle,decoration.filled=30}{v1}
\fmf{plain}{v1,vc1}
\fmf{plain}{vc1,v2}
\fmf{plain}{v1,vc2}
\fmf{plain}{vc2,v3}
\fmffreeze
\fmf{photon,left=0.45}{vc1,vc2}
\fmf{photon,right=0.45}{vc1,vc2}
\fmffreeze
\fmfposition
\end{fmfgraph*}}
&
\parbox{20mm}{
\begin{fmfgraph*}(17,40)
\fmftop{v1}
\fmfleft{v2}
\fmfright{v3}
\fmfv{decoration.shape=circle,decoration.filled=30}{v1}
\fmf{plain}{v1,vc1}
\fmf{plain}{vc1,vc3}
\fmf{plain}{vc3,v2}
\fmf{plain}{v1,vc2}
\fmf{plain}{vc2,v3}
\fmffreeze
\fmf{photon,left=0.15}{vc1,vc2}
\fmf{photon,right=0.15}{vc3,vc2}
\fmffreeze
\fmfposition
\end{fmfgraph*}}
&
\parbox{20mm}{
\begin{fmfgraph*}(17,40)
\fmftop{v1}
\fmfleft{v2}
\fmfright{v3}
\fmfv{decoration.shape=circle,decoration.filled=30}{v1}
\fmf{plain}{v1,vc1}
\fmf{plain}{vc1,vc3}
\fmf{plain}{vc3,vc4}
\fmf{plain}{vc4,v2}
\fmf{plain}{v1,vc2}
\fmf{plain}{vc2,v3}
\fmffreeze
\fmf{photon,right=0.65}{vc1,vc4}
\fmf{photon,right=0.15}{vc3,vc2}
\fmffreeze
\fmfposition
\end{fmfgraph*}}
&
\parbox{20mm}{
\begin{fmfgraph*}(17,40)
\fmftop{v1}
\fmfleft{v2}
\fmfright{v3}
\fmfv{decoration.shape=circle,decoration.filled=30}{v1}
\fmf{plain}{v1,vc1}
\fmf{plain}{vc1,vc4}
\fmf{plain}{vc4,v2}
\fmf{plain}{v1,vc2}
\fmf{plain}{vc2,v3}
\fmffreeze
\fmf{photon,left=0.25}{vc1,vc5}
\fmf{photon,right=0.25}{vc4,vc5}
\fmf{photon,right=0.05}{vc5,vc2}
\fmffreeze
\fmfposition
\end{fmfgraph*}}\\
\parbox{20mm}{\vspace{-3cm}$~~\quad I_1$} 
& 
\parbox{20mm}{\vspace{-3cm}$~~\quad I_2$} 
&
\parbox{20mm}{\vspace{-3cm}$~~\quad I_3$} 
&
\parbox{20mm}{\vspace{-3cm}$~~\quad I_4$} \\
\parbox{20mm}{
\begin{fmfgraph*}(17,40)
\fmftop{v1}
\fmfleft{v2}
\fmfright{v3}
\fmfv{decoration.shape=circle,decoration.filled=30}{v1}
\fmf{plain}{v1,vc1}
\fmf{plain}{vc1,v2}
\fmf{plain}{v1,vc2}
\fmf{plain}{vc2,v3}
\fmffreeze
\fmf{photon,right=0.15}{vc1,vc3}
\fmf{photon,right=0.15}{vc3,vc2}
\fmfv{d.sh=circle,l.d=0, d.f=empty,d.si=.3w,l=$1$}{vc3}
\fmffreeze
\fmfposition
\end{fmfgraph*}}
&
\parbox{20mm}{
\begin{fmfgraph*}(17,40)
\fmftop{v1}
\fmfleft{v2}
\fmfright{v3}
\fmfv{decoration.shape=circle,decoration.filled=30}{v1}
\fmf{plain,left=0.7}{v1,vc1}
\fmf{plain,right=0.7}{v1,vc1}
\fmf{dashes,left=0.7}{vc1,vc2}
\fmf{dashes,right=0.7}{vc1,vc2}
\fmf{plain}{vc2,v2}
\fmf{plain}{vc2,v3}
\end{fmfgraph*}}
&
\parbox{20mm}{
\begin{fmfgraph*}(17,40)
\fmftop{v1}
\fmfleft{v2}
\fmfright{v3}
\fmfv{decoration.shape=circle,decoration.filled=30}{v1}
\fmf{plain}{v1,vc1}
\fmf{plain}{vc1,v2}
\fmf{plain}{v1,vc2}
\fmf{plain}{vc2,v3}
\fmffreeze
\fmf{dashes,left=0.45}{vc1,vc2}
\fmf{dashes,right=0.45}{vc1,vc2}
\fmffreeze
\fmfposition
\end{fmfgraph*}}
&
\parbox{20mm}{
\begin{fmfgraph*}(17,40)
\fmftop{v1}
\fmfleft{v2}
\fmfright{v3}
\fmfv{decoration.shape=circle,decoration.filled=30}{v1}
\fmf{plain}{v1,vc1}
\fmf{plain}{vc1,vc3}
\fmf{plain}{vc3,v2}
\fmf{plain}{v1,vc2}
\fmf{plain}{vc2,vc4}
\fmf{plain}{vc4,v3}
\fmffreeze
\fmf{photon,right=0.15}{vc1,vc4}
\fmf{photon,right=0.15}{vc3,vc2}
\fmffreeze
\fmfposition
\end{fmfgraph*}}\\
\parbox{20mm}{\vspace{-3cm}$~~\quad I_5$} 
& 
\parbox{20mm}{\vspace{-3cm}$~~\quad I_6$} 
&
\parbox{20mm}{\vspace{-3cm}$~~\quad I_7$} 
&
\parbox{20mm}{\vspace{-3cm}$~~\quad I_8$}
\end{tabular}
\end{center}
\vspace{-1cm}
These diagrams also appear in \cite{Minahan:2009wg}, where effective
Feynman rules for them have been given. These rules are reproduced
here in appendix \ref{sec:app}, along with supplementary details of
the calculation, action, conventions etc. for convenience. The
effective Feynman rules come from replacing momentum contractions
involving the three-dimensional Levi-Civita tensor with usual scalar
(i.e. dot) products. We then use these scalar products to remove
propagators from a scalar ``master topology'' diagram; the results are
then reduced to master integrals using the Laporta algorithm
\cite{Laporta:2001dd}. It may seem surprising to see in $I_8$ a
non-planar-looking diagram\footnote{We have verified that the
  ``un-crossed'' version of $I_8$ is exactly zero; an argument for it
  being ${\cal O}(\e)$ or smaller is given in appendix
  \ref{sec:appv}.}, even though we are working in the large-$N$ limit:
since one of the gluons may travel around the outside of the operator,
the diagram is in fact planar. The colour structure will be discussed
in greater detail below. Diagram $I_6$ is trivial to evaluate and we
will consider it later. Apart from $I_8$ which requires a different
treatment, the remaining diagrams may all be expressed in terms of the
following master topology
\vspace{0.4cm}
\begin{center}\hspace{-4cm}
\parbox{20mm}{
\begin{fmfgraph*}(65,21)
\fmfleft{i1,i2}
\fmflabel{$p_2$}{i1}
\fmflabel{$p_1+p_2$}{i2}
\fmfright{o1,o2}
\fmflabel{$p_2$}{o1}
\fmflabel{$p_1+p_2$}{o2}
\fmf{fermion}{i1,v1}
\fmf{fermion,tension=.5,label=$q$,l.side=right}{v1,v3a}
\fmf{fermion,tension=.5,label=$l$,l.side=right}{v3a,v3}
\fmf{fermion}{v3,o1}
\fmf{fermion}{o2,v4}
\fmf{fermion,tension=.5,label=$l+p_1$,l.side=right}{v4,v2a}
\fmf{fermion,tension=.5,label=$q+p_1$,l.side=right}{v2a,v2}
\fmf{fermion}{v2,i2}
\fmf{fermion,tension=0,label=$q-l$}{v3a,v2a}
\fmf{fermion,tension=.2,label=$q-p_2$,l.side=right}{v2,v1}
\fmf{fermion,tension=.2,label=$l-p_2$,l.side=right}{v3,v4}
\end{fmfgraph*}}
\end{center}\vspace{0.5cm}

\ni where we label propagators as follows
\vspace{0.5cm}
\begin{center}\hspace{-4cm}
\parbox{20mm}{
\begin{fmfgraph*}(65,25)
\fmfleft{i1,i2}
\fmfright{o1,o2}
\fmf{plain}{i1,v1}
\fmf{plain,tension=.5,label=$2$,l.side=right}{v1,v3a}
\fmf{plain,tension=.5,label=$6$,l.side=right}{v3a,v3}
\fmf{plain}{v3,o1}
\fmf{plain}{v4,o2}
\fmf{plain,tension=.5,label=$7$,l.side=left}{v2a,v4}
\fmf{plain,tension=.5,label=$3$,l.side=right}{v2a,v2}
\fmf{plain}{i2,v2}
\fmf{plain,tension=0,label=$4$}{v3a,v2a}
\fmf{plain,tension=.2,label=$1$,l.side=right}{v2,v1}
\fmf{plain,tension=.2,label=$5$,l.side=left}{v4,v3}
\end{fmfgraph*}}
\vspace{1cm}
\end{center}
\ni This allows us to express all terms in the Feynman rules in terms
of the integral
\bsp\label{G7}
G(n_1,n_2,n_3,n_4,n_5,n_6,n_7) = &\int \frac{d^{2\o}q}{(2\pi)^{2\o}} 
\int \frac{d^{2\o}l}{(2\pi)^{2\o}} 
\left[(q-p_2)^2\right]^{-n_1}
 \left[q^2\right]^{-n_2}
\left[(q+p_1)^2\right]^{-n_3}\\
&\left[(q-l)^2\right]^{-n_4}
\left[(l-p_2)^2\right]^{-n_5}
\left[l^2\right]^{-n_6}
\left[(l+p_1)^2\right]^{-n_7}.
\end{split}
\ee
Take for example $I_1$, this diagram has $n_1=n_3=n_6=0$ prior to
consideration of the numerator. Using the first rule in appendix
\ref{sec:app} \cite{Minahan:2009wg} we see that the numerator is a
scalar product of the two momenta (those labeled by 2 and 4 in the
master topology) carried by the two gluons in the diagram. We
reexpress this scalar product in terms of $(p_1+p_2)^2$ and the
squared momentum configurations found in the propagators of the master
topology
\be
-\frac{1}{2}\,q\cdot (q-l) = \frac{1}{4}\left(-q^2-(q-l)^2  + l^2\right),  
\ee
or, in terms of the $G$ integrals
\bsp
I_1 = \frac{1}{4} \Bigl( -G(0,0,0,1,1,0,1)-
G(0,1,0,0,1,0,1) + G(0,1,0,1,1,-1,1) \Bigr),
\end{split}
\ee
where colour information has been stripped; this will be restored
later. Another example is
\bsp
I_5 = 
&G(0, 1, 0, 1, 0, 0, 1) - G(0, 1, 0, 1, 0, 1, 0) - 
 G(0, 1, 0, 1, 1, -1, 1)\\ + &G(0, 1, 0, 1, 1, 0, 0) - 
 2\, s\, G(0, 1, 0, 1, 1, 0, 1)
\end{split}
\ee
where $s=(p_1+p_2)^2$. 

We may then use the Laporta algorithm \cite{Laporta:2001dd} and reduce
these expressions to master integrals. We have used the software
package FIRE \cite{Smirnov:2008iw} to achieve this. The results are
\bsp
&I_1 = -\frac{s}{4} \frac{2\o-3}{3\o-4}\, F_1,\quad
I_2 = \frac{1}{4} \left( SS +\frac{3s}{2}  \frac{2\o-3}{3\o-4}\, F_1
\right),\\
&I_3 = \frac{1}{2} (4\o-5) \left(\frac{1}{2\o-3}\, SS +
\frac{s}{3\o-4} \,F_1 \right),\\
&I_4 = \frac{1}{2}I_3,\quad
I_5 = -\frac{s\,(4\o-5)}{3\o-4}\,F_1,\\
&I_6 = -\frac{1}{(4\pi)^{2\o}}\frac{1}{s^{3-2\o}}\left(
\frac{\G(2-\o)\G^2(\o-1)}{\G(2\o-2)}\right)^2,\\
&I_7 = -4 \,I_1,
\end{split}
\ee
where the master integrals $SS=G(0,0,1,1,1,0,0)$ and
$F_1=G(0,1,0,1,1,0,1)$ may be visualized as follows 
\vspace{-1cm}
\[F_1 = 
\parbox{20mm}{\vspace{2cm}
\begin{fmfgraph*}(17,40)
\fmftop{v1}
\fmfleft{v2}
\fmfright{v3}
\fmf{plain}{v1,v2}
\fmf{plain}{v1,v3}
\fmffreeze
\fmf{plain,left=0.45}{v2,v3}
\fmf{plain,right=0.45}{v2,v3}
\fmffreeze
\fmfposition
\end{fmfgraph*}},\qquad
SS=
\parbox{20mm}{\vspace{0cm}
\begin{fmfgraph*}(17,20)
\fmftop{v1}
\fmfbottom{v2}
\fmf{plain}{v1,v2}
\fmffreeze
\fmf{plain,left=0.45}{v1,v2}
\fmf{plain,right=0.45}{v1,v2}
\fmffreeze
\fmfposition
\end{fmfgraph*}}.\vspace{-1cm}\]
These integrals may be calculated in closed form by standard
techniques
\bsp
&F_1 = \frac{1}{(4\pi)^{2\o}} \frac{1}{s^{4-2\o}}
\frac{\G(4-2\o)}{\G(3\o-4)}
\G(2-\o)\G^2(\o-1)\frac{\G(2\o-3)}{2\o-3},\\
&SS = \frac{1}{(4\pi)^{2\o}} \frac{1}{s^{3-2\o}}
\frac{\G(3-2\o)\G^3(\o-1)}{\G(3\o-3)}.
\end{split}
\ee

A crossed topology is required to reduce $I_8$, see appendix
\ref{sec:appi8}. We use the following
\vspace{1cm}
\begin{center}\hspace{-4cm}
\parbox{20mm}{
\begin{fmfgraph*}(45,35)
\fmfleft{i1,i2}
\fmflabel{$p_1$}{i1}
\fmflabel{$p_1$}{i2}
\fmfright{o1,o2}
\fmflabel{$p_2$}{o1}
\fmflabel{$p_2$}{o2}
\fmf{fermion,tension=0.4}{v1,i1}
\fmf{fermion,tension=.01,label=$l$,l.side=right,label.dist=-20}{v1,v4}
\fmf{fermion,tension=0.4}{v3,o1}
\fmf{fermion}{o2,v4c}
\fmf{fermion,tension=.2,label=$q$,l.side=right}{v4c,v2}
\fmf{fermion}{i2,v2}
\fmf{fermion,tension=.2,label=$q+p_1$,l.side=right}{v2,v1c}
\fmf{fermion,tension=.2,label=$q-l+p_1$,l.side=right}{v1c,v1}
\fmf{fermion,tension=.2,label=$l-p_2$,l.side=right}{v3,v4}
\fmf{fermion,tension=.2,label=$q-p_2$,l.side=right}{v4,v4c}
\fmf{fermion,tension=.01,label=$q-l$,l.side=left,label.dist=16}{v1c,v3}
\end{fmfgraph*}}
\hspace{4cm}
\parbox{20mm}{
\begin{fmfgraph*}(55,35)
\fmfleft{i1,i2}
\fmfright{o1,o2}
\fmf{plain}{v1,i1}
\fmf{plain,tension=.01,label=$4$,l.side=right,label.dist=-17}{v1,v4c}
\fmf{plain}{v3,o1}
\fmf{plain}{o2,v4}
\fmf{plain,tension=.2,label=$1$,l.side=right}{v4,v2}
\fmf{plain}{i2,v2}
\fmf{plain,tension=.2,label=$2$,l.side=right}{v2,v1c}
\fmf{plain,tension=.2,label=$6$,l.side=right}{v1c,v1}
\fmf{plain,tension=.2,label=$7$,l.side=right}{v3,v4c}
\fmf{plain,tension=.2,label=$3$,l.side=right}{v4c,v4}
\fmf{plain,tension=.01,label=$5$,l.side=left,label.dist=10}{v1c,v3}
\end{fmfgraph*}}
\end{center}\vspace{1cm}
and so associate an integral $G(n_1,\ldots,n_7)$ analogous to (\ref{G7}). We obtain
\bsp
I_8 = &-\frac{s^3}{4}\frac{\o-2}{4\o-7} \,X_1
+\frac{1}{8}\left(43+\frac{12}{3-2\o}+\frac{28}{(\o-2)^2}
+\frac{70}{\o-2}+\frac{35}{4\o-7}\right)\,SS\\
&+\frac{s}{24}\left(-37+\frac{16}{4-3\o}-\frac{30}{\o-2}
+\frac{15}{4\o-7}\right)\,F_1,
\end{split}
\ee
where $X_1 = G(0, 1, 1, 1, 1, 1, 1)$ in this topology and may be
visualized as follows
\begin{equation*}
X_1 = 
\parbox{20mm}{\vspace{1cm}
\begin{fmfgraph*}(35,35)
\fmftop{i1}
\fmfbottom{o1,o2}
\fmf{plain}{i1,v1}
\fmf{plain}{v1,v2}
\fmf{plain,tension=5}{v2,o1}
\fmf{plain}{i1,v3}
\fmf{plain}{v3,v4}
\fmf{plain,tension=5}{v4,o2}
\fmf{plain,tension=.01}{v1,v4}
\fmf{plain,tension=.01}{v2,v3}
\end{fmfgraph*}}~~~~~~~~.
\end{equation*}
It is given by \cite{Gehrmann:2005pd}
\bsp
&X_1 = \frac{1}{2s^{6-2\o}}
\frac{1}{(4\pi)^{2\o}}\frac{\G(5-2\o)\G(\o-1)}{(\o-2)^4}
\Biggl(\\
& -\frac{\G^2(\o-1)}{\G(3\o-5)}\,
      {}_4F_3(1,\o-1,2\o-4,4\o-8;2\o-3,2\o-3,3\o-5;1)\\
&-8\frac{(\o-2)^2\G(\o-1)\G(2\o-3)}{(\o-3)(2\o-5)\G(4\o-7)}\,
{}_3F_2(1,1,5-2\o;6-2\o,4-\o;1)\\
&+\frac{\G(3-\o)\G(\o-1)\G(2\o-3)}{\G(3\o-5)}\,
{}_3F_2(1,2\o-4,4\o-8;2\o-3,3\o-5;1)\\
&-2\frac{\G^2(5-2\o)\G(3-\o)\G^4(2\o-3)}{\G(9-4\o)\G^2(4\o-7)}\Biggr).
\end{split}
\ee
%

\subsubsection{Colour factors}

The colour structure of the ABJ(M) fields is as follows: Chern-Simons
gauge fields are $U(N)$ adjoints $A^\m_{ij}$ and $U(M)$ adjoints $\hat
A^\m_{\hat i \hat j}$ (with opposite-sign Chern-Simons level), where
$i,j=1,\ldots,N$ and $\hat i,\hat j=1,\ldots,M$. In the ABJM case
$N=M$, while ABJ is defined by $N\neq M$. There are complex scalars
with flavour group $SU(4)$ ($A$ is a fundamental $SU(4)$ index) which
transform in the bifundamental $(N,\bar N)$ of the two $U(N)$ groups,
i.e. $(Y^A)_{i\hat i}$ and their anti-bifundamental $(\bar N, N)$
conjugates $(Y^\dag_A)_{\hat i i}$. There are accompanying complex
two-component fermions $(\psi_A)_{i \hat i}$ and $(\psi^{\dag
  A})_{\hat i i}$. The action for the ABJ(M) theory is given in (\ref{ABJac}).

\begin{figure}
\begin{center}
\includegraphics[bb=35 35 265 265,width=1.5in,clip=true]{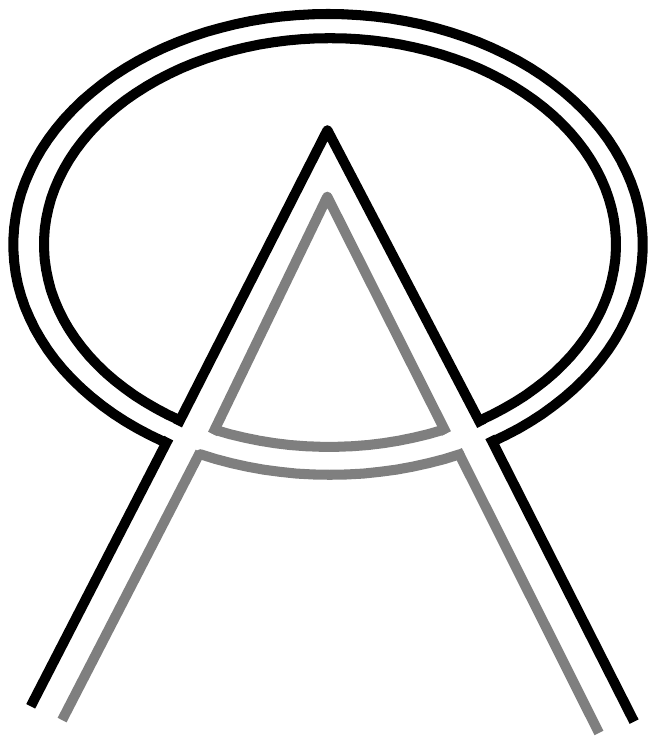}
\includegraphics[bb=35 35 275 265,width=1.5in,clip=true]{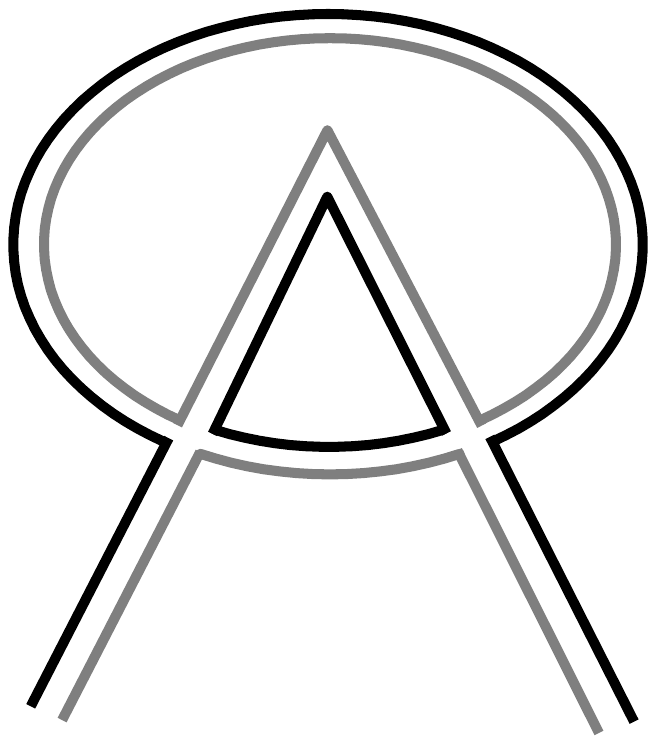}
\includegraphics[bb=70 35 190 240,width=0.9in,clip=true]{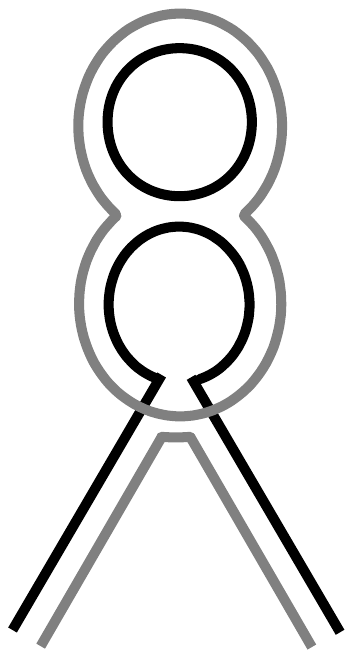}
\end{center}
\caption{We show fat graphs for would-be-non-planar contributions to
  $I_1$ (shown on the left), $I_7$ (shown in the center), and $I_6$
  (shown on the right) where unhatted index lines are represented by a
  black line while hatted index lines are coloured gray. Note that
  $I_2$ and $I_8$ have topologies identical to $I_1$.}\label{fig:np}
\end{figure}

There are certain contributions to the $J=2$ case which are special,
in that their analogues for the $J> 2$ cases are non-planar. These
are versions of $I_1$, $I_2$, $I_6$, $I_7$, and\footnote{$I_7$ and
  $I_8$ have no normally planar version, and thus only contribute to the
  $J=2$ case.} $I_8$ where one of the exchanged particles must go
``around the outside'' of the graph, and therefore in the case that
there are more than two external legs, would have to cross them and
thus become non-planar. The fat graphs for these contributions are
given in figure \ref{fig:np}.

In the case of $I_1$ and $I_2$ the net effect of adding-in these
would-be-non-planar contributions is a sign flip. The accounting for
this is simple to see; the would-be-non-planar graphs involve one
gluon of each flavour and therefore arise from cross-terms in the
$D_\m Y^A D^\m Y^\dag_A$ part of the action. This means both a factor
of two and a sign, relative to the contributions involving the same
flavour of gluon, thus the net effect is $1-2=-1$, or a flipped sign.

In the case of $I_7$ there is a non-trivial flavour structure at
play. The fermion-scalar vertices come in two varieties
\bsp
&V_1 =\Tr\left( Y^A Y^\dag_B \psi_C \psi^{\dag D} \right)
\left(\d^B_A\d^C_D - 2 \d^B_D\d^C_A\right),\\
&V_2 = -\Tr\left( Y^A \psi^{\dag B} Y^C \psi^{\dag D}\right) \e_{ABCD} ,
\end{split}
\ee
and (minus) their conjugates $-\bar V_1$ and $-\bar V_2$. It is clear
that we cannot take one of each variety in the graphs of interest to
us, because the fields only propagate to their conjugates. For the
diagram $I_7$ the contributions involving the $V_1$-$V_1$, $\bar
V_1$-$\bar V_1$, and $V_1$-$\bar V_1$ vertex combinations are
proportional to the flavour trace, and so vanish, whereas the
$V_2$-$\bar V_2$ combination involves a contraction of the two epsilon
tensors on two indices, leaving both a flavour trace (which is zero)
and an identity operator, which is not.

The case of $I_6$ is different in that the would-be-non-planar
contributions from the $V_1$-$\bar V_1$ vertex combination (see figure
\ref{fig:np}) are in fact equal and opposite to the normally planar
contributions from the $V_1$-$V_1$ and $\bar V_1$-$\bar V_1$ vertex
combinations. This cancellation eliminates the diagram $I_6$ in the
$J=2$ case.

The diagrams $I_7$ and $I_8$ have no normally planar counter-parts and
thus only appear for $J=2$.

\subsubsection{Assembling the result}

We now assemble the result of our calculation. Diagrams $I_2$, $I_3$,
and $I_4$ receive a factor of two because we must add the left $\lr$
right versions of them. As explained above the would-be-non-planar
contributions flip the sign of $I_1$ ($I_7$ earns a similar minus sign)
and $I_2$, and eliminate $I_6$. We therefore have that
\be\label{ffres}
F_{ABJM}(s) = 2\,(4\pi\l)^2\,
\Bigl(-I_1 -2\,I_2 +2\,I_3 +2\,I_4+I_5+0\cdot I_6-I_7+ I_8\Bigr),
\ee 
where $\l =  N/k$ is the 't Hooft coupling and where the leading
factor of two counts the two $U(N)$ gauge groups, or equivalently the
sum over the even and odd ``sites'' in the CPO.

Expanding (using the HypExp package \cite{Huber:2005yg}) 
in $d=2\o = 3-2\e$, we find 
\be\label{rJ2ABJM}
F_{ABJM}(s) = \frac{\l^2}{2} \left(\frac{s\,e^\g }{4\pi} \right)^{-2\e} \left(
-\frac{1}{2\e^2} -\frac{\log 2}{\e} +2\log^2
2+\frac{\pi^2}{3}
+{\cal O}(\e)\right).
\ee

It is also a straight-forward exercise to promote this result to the
ABJ case, where the two gauge groups have different ranks. We define
an extra 't Hooft coupling $\hat\l = M/k$, where $M$ is the rank of
one of the gauge groups, while the other remains $N$. By counting the
light and dark gray index loops in diagrams like those shown in figure
\ref{fig:np} we find\footnote{I thank Marco S. Bianchi, Marta Leoni,
  Matias Leoni, Andrea Mauri, Silvia Penati, and Alberto Santambrogio for
  pointing out an error in the colour factor for $I_6$ present in an
  earlier version of this manuscipt.}
\bsp
F_{ABJ}(s) = &\left(\frac{4\pi}{k}\right)^2\,
\Bigl((M^2+N^2-4MN)(I_1 +2\,I_2) \\
&+2MN(2I_3 +I_5-I_7+ I_8)
+ 2\,(M^2+N^2)\,I_4+(M-N)^2\, I_6\Bigr),
\end{split}
\ee 
which upon expansion yields
\bsp\label{rJ2ABJ}
F_{ABJ}(s) = \frac{1}{2} \left(\frac{s\,e^\g }{4\pi} \right)^{-2\e} \Biggl(
&-\frac{\l\hat\l}{2\e^2} -(\l^2+\hat\l^2)\frac{\log 2}{2\e}
 +(\l^2+\hat\l^2)\log^2 2\\
&-\Bigl( 11\l^2+11\hat\l^2-30\l\hat\l\Bigr)\, \frac{\pi^2}{24}
+{\cal O}(\e)\Biggr).
\end{split}
\ee
We note that both the ABJM result in (\ref{rJ2ABJM}) and the ABJ
result in (\ref{rJ2ABJ}) respect the principle of maximal transcendentality.

\subsection{$J> 2$ case}

Here we will encounter two new diagrams, shown below for the case
$J=3$ ($J$ must be even, this is just for visualization
purposes)\footnote{The diagram corresponding to two separate gluon
  exchanges between legs 1 and 2, and legs 2 and 3 vanishes, see
  appendix \ref{sec:appv}.}
\begin{center}
\begin{tabular}[t]{cc}
\parbox{20mm}{\vspace{1cm}
\begin{fmfgraph*}(17,20)
\fmftop{v1}
\fmfstraight
\fmfbottom{v2,v3,v4}
\fmfv{decoration.shape=circle,decoration.filled=30}{v1}
\fmf{plain}{v1,vc1}
\fmf{plain}{vc1,v2}
\fmf{plain}{v1,vc2}
\fmf{plain}{vc2,v3}
\fmf{plain}{v1,vc3}
\fmf{plain}{vc3,v4}
\fmffreeze
\fmf{photon,right=0.15}{vc1,vc2}
\fmf{photon,right=0.15}{vc2,vc3}
\end{fmfgraph*}\vspace{0.5cm}}
&\hspace{1cm}
\parbox{20mm}{\vspace{1cm}
\begin{fmfgraph*}(17,20)
\fmftop{v1}
\fmfstraight
\fmfbottom{v2,v3,v4}
\fmfv{decoration.shape=circle,decoration.filled=30}{v1}
\fmf{plain,left}{v1,vc1}
\fmf{plain}{vc1,v2}
\fmf{plain}{v1,vc1}
\fmf{plain}{vc1,v3}
\fmf{plain,right}{v1,vc1}
\fmf{plain}{vc1,v4}
\end{fmfgraph*}\vspace{0.5cm}}\\
$I_9$ &\hspace{1cm} $I_{10}$
\end{tabular}
\end{center}

\ni The master topology we need for these graphs is

\vspace{0.5cm}
\begin{center}\hspace{-4cm}
\parbox{20mm}{
\begin{fmfgraph*}(75,35)
\fmfcurved
\fmfleft{i1,i2,i3}
\fmflabel{$p_1$}{i1}
\fmflabel{$p_1+2\,p_2+p_3$}{i2}
\fmflabel{$p_2$}{i3}
\fmfright{o1,o2,o3}
\fmflabel{$p_1+p_2+p_3$}{o1}
\fmflabel{$p_1+p_3$}{o2}
\fmflabel{$p_3$}{o3}
\fmf{fermion}{v1,i1}
\fmf{fermion,tension=.1}{i2,v2}
\fmf{fermion}{v3,i3}
\fmf{fermion,tension=.5,label=$k+p_2+p_3$,l.side=right}{v1,cb}
\fmf{fermion,tension=.5,label=$l+p_2+p_3$,l.side=right}{cb,l1}
\fmf{fermion}{l1,o1}
\fmf{fermion,tension=.1}{o2,l2}
\fmf{fermion}{l3,o3}
\fmf{fermion,tension=.5,label=$l$,l.side=right}{l3,ct}
\fmf{fermion,tension=.5,label=$k$,l.side=right}{ct,v3}
\fmf{fermion,tension=0,label=$l-k$}{ct,cb}
\fmf{fermion,tension=.75,label=$k+p_1+p_2+p_3$,l.side=right}{v2,v1}
\fmf{fermion,tension=.75,label=$k-p_2$,l.side=right}{v3,v2}
\fmf{fermion,tension=.75,label=$l-p_1$,l.side=right}{l1,l2}
\fmf{fermion,tension=.75,label=$l+p_3$,l.side=right}{l2,l3}
\end{fmfgraph*}}
\vspace{1cm}
\end{center}
\ni where we label propagators as
\vspace{1cm}
\begin{center}\hspace{-4cm}
\parbox{20mm}{
\begin{fmfgraph*}(75,35)
\fmfcurved
\fmfleft{i1,i2,i3}
\fmfright{o1,o2,o3}
\fmf{plain}{v1,i1}
\fmf{plain,tension=.1}{i2,v2}
\fmf{plain}{v3,i3}
\fmf{plain,tension=.5,label=$4$,l.side=right}{v1,cb}
\fmf{plain,tension=.5,label=$6$,l.side=right}{cb,l1}
\fmf{plain}{l1,o1}
\fmf{plain,tension=.1}{o2,l2}
\fmf{plain}{l3,o3}
\fmf{plain,tension=.5,label=$9$,l.side=right}{l3,ct}
\fmf{plain,tension=.5,label=$1$,l.side=right}{ct,v3}
\fmf{plain,tension=0,label=$5$}{ct,cb}
\fmf{plain,tension=.75,label=$3$,l.side=right}{v2,v1}
\fmf{plain,tension=.75,label=$2$,l.side=right}{v3,v2}
\fmf{plain,tension=.75,label=$7$,l.side=right}{l1,l2}
\fmf{plain,tension=.75,label=$8$,l.side=right}{l2,l3}
\end{fmfgraph*}}
\end{center}\vspace{1cm}
and associate a two-loop integral analogous to (\ref{G7}) which we
call $G(n_1,\ldots,n_9)$. Using the effective
rule of eq. (A.7) of \cite{Minahan:2009wg} (reproduced here in appendix \ref{sec:app}),
and the FIRE package, we find the following result
\bsp
I_9 = -\frac{1}{4} \frac{1}{(2\o-3)(3\o-5)}\Biggl[
&(50+2\o(12\o-35)) \Bigl(SS(s_{23}) - SS(s_{12}+s_{13}+s_{23}) +
SS(s_{12}) \Bigr)\\
&+(2\o-4)^2 s_{12}\, s_{23} \, F_3 \Biggr],
\end{split}
\ee
where $F_3 = G(0,0,1,1,1,0,0,1,1)$ and may be visualized by the
following figure
\begin{equation*}
F_3 = 
\parbox{20mm}{
\begin{fmfgraph*}(17,20)
\fmfstraight
\fmftop{v1}
\fmfbottom{v2,v3,v4}
\fmf{plain}{v1,v2}
\fmf{plain}{v1,v3}
\fmf{plain}{v1,v4}
\fmf{plain}{v2,v3}
\fmf{plain}{v2,v4}
\end{fmfgraph*}}.
\end{equation*}
This integral has been provided in eq. (5.27) of
\cite{Gehrmann:1999as}. The result is
\bsp\label{F3}
&F_3 = \frac{(3\o-4)(3\o-5)}{(\o-2)^2} \frac{SS(1)}{s_{12}s_{23}}\\
\times\Biggl[&-\left(\frac{s_{12}s_{23}}{s_{13}+s_{23}}\right)^{2\o-3}
{}_2F_1\left(2\o-3,2\o-3,2\o-2,\frac{s_{13}}{s_{13}+s_{23}}\right)\\
&-\left(\frac{s_{12}s_{23}}{s_{13}+s_{12}}\right)^{2\o-3}
{}_2 F_1\left(2\o-3,2\o-3,2\o-2,\frac{s_{13}}{s_{13}+s_{12}}\right)\\
&+\left(\frac{s_{123}s_{12}s_{23}}{(s_{13}+s_{23})(s_{13}+s_{12})}\right)
^{2\o-3}
{}_2F_1\left(2\o-3,2\o-3,2\o-2,\frac{s_{13}s_{123}}{(s_{13}+s_{12})(s_{13}+s_{23})}\right)\Biggr],
\end{split}
\ee
where $s_{123} = s_{12}+s_{23}+s_{13}$. The diagram $I_{10}$ is
trivial
\be
I_{10} = SS(s_{12}+s_{13}+s_{23}).
\ee
%

\subsubsection{Flavour structure for $I_{10}$}

\ni The flavour structure of the six-vertex gives (see eq. (B.2) of
\cite{Minahan:2009wg})
\be
{\bf 1} - \frac{1}{2} P
\ee
where $P$ permutes two nearest odd or even sites. Since the CPO is
symmetric, this gives rise to an extra factor of $1/2$ dressing
$I_{10}$, see below.

\subsubsection{Assembling the result}

\ni We must note that now that we have more than two legs, the
diagrams $I_7$ and $I_8$ do not contribute at the planar level, $I_6$
is not canceled-out, and there are not sign-flips on $I_1$ and
$I_2$. Thus we have that
\bsp
F_{ABJM}(\{s_{ij}\}) = (4\pi\l)^2 \sum_{\substack{\text{even and odd sites}\\ \text{$\equiv$ legs}}}
\Bigl( I_1 + 2\left(I_2+I_3+I_4\right) + I_5 + I_6 + I_9 +\frac{1}{2} I_{10}\Bigr).
\end{split}
\ee
Expanding in $\e$, we find that
\bsp\label{rJJABJM}
F_{ABJM}(\{s_{ij}\}) = &\frac{J\l^2}{4} \left(\frac{e^\g }{4\pi} \right)^{-2\e}\Biggl(
-\frac{1}{2\e^2} +\frac{1}{\e}\frac{1}{J}\sum_{i=1}^J \log\frac{s_{ii+1}}{2}\\
&+2\log^2 2 -\frac{\pi^2}{3} - \frac{1}{J}\sum_{i=1}^J\log\frac{s_{ii+1}}{4}\log s_{ii+1}
-\text{Trans}_2(\{s_{ij}\})
+{\cal O}(\e)\Biggr),
\end{split}
\ee
where 
\bsp
\text{Trans}_2(\{s_{ij}\}) = &\frac{1}{J}\sum_{i=1}^J\Biggl(
\log s_{ii+1}\log s_{i+1i+2}\\
&-\log(s_{ii+1}+s_{ii+2})\log(s_{ii+2}+s_{i+1i+2})\\
&+\log\frac{(s_{ii+1}+s_{ii+2})(s_{ii+2}+s_{i+1i+2})}{s_{ii+1}s_{i+1i+2}}
\log(s_{ii+1}+s_{i+1i+2}+s_{ii+2})\\
&+\text{Li}_2\frac{s_{ii+2}}{s_{ii+1}+s_{ii+2}}
+\text{Li}_2\frac{s_{ii+2}}{s_{i+1i+2}+s_{ii+2}}\\
&+\text{Li}_2\frac{(s_{ii+1}+s_{i+1i+2}+s_{ii+2})s_{ii+2}}
{(s_{ii+1}+s_{ii+2})(s_{ii+2}+s_{i+1i+2})}
\Biggr).
\end{split}
\ee

The ABJ case can be similarly worked out, we find
\bsp
F_{ABJ}(\{s_{ij}\}) = \left(\frac{4\pi}{k}\right)^2 \Biggl[
&\sum_{\text{odd sites}}
\Bigl( M^2 I_1 + 2\left(M^2 I_2+ MN I_3+ M^2 I_4\right) + MNI_5\\
& \qquad\qquad\qquad+ M^2I_6 + MNI_9 +\frac{MN}{2} I_{10}\Bigr)\\
+&\sum_{\text{even sites}}
\Bigl( N^2 I_1 + 2\left(N^2 I_2+ MN I_3+ N^2 I_4\right) + MNI_5\\
& \qquad\qquad\qquad+ N^2I_6 + MNI_9 +\frac{MN}{2} I_{10}\Bigr)\Biggr],
\end{split}
\ee
which expands to give
\bsp\label{rJJABJ}
F_{ABJ}(\{s_{ij}\}) = &\frac{J}{4} \left(\frac{e^\g }{4\pi} \right)^{-2\e}\Biggl(
-\frac{\l\hat\l}{2\e^2} -\frac{\l^2+\hat\l^2}{2\e}\log 2 
+\frac{\l\hat\l}{\e}\frac{1}{J}\sum_{i=1}^J \log s_{ii+1}\\
&+(\l^2+\hat\l^2)\log^2 2
-\Bigl(11(\l^2+\hat\l^2)-14\l\hat\l\Bigr)\frac{\pi^2}{24}\\
& +(\l^2+\hat\l^2) \frac{1}{J}\sum_{i=1}^J\log 2\,\log s_{ii+1}
-\l\hat\l \frac{1}{J}\sum_{i=1}^J\log^2 s_{ii+1}\\
&-\l\hat\l\,\text{Trans}_2(\{s_{ij}\})
+{\cal O}(\e)\Biggr).
\end{split}
\ee
We note that, as in the $J=2$ case, both the ABJM result in
(\ref{rJJABJM}) and the ABJ result in (\ref{rJJABJ}) respect the
principle of maximal transcendentality.

\section{Discussion}
\label{sec:disc}

We have computed the leading quantum correction to the form factors
for scalar chiral primary operators in ABJ(M) to an equal number of
scalar final states at leading order in the `t Hooft coupling. Our
results, given in (\ref{rJ2ABJM}), (\ref{rJ2ABJ}), (\ref{rJJABJM}),
and (\ref{rJJABJ}) are seen to obey the principle of maximal
transcendentality, in that the terms of ${\cal O}(\e^{-n})$ are of
transcendentality $2-n$. This is consistent with calculations both of
scattering amplitudes
\cite{Agarwal:2008pu,Chen:2011vv,Bargheer:2012cp,Bianchi:2012cq,CaronHuot:2012hr}
and, recently, with light-like Wilson loop computations
\cite{Bianchi:2013pva} (see
\cite{Henn:2010ps,Bianchi:2011rn,Bianchi:2011dg,Wiegandt:2011uu} for
previous work) in the same theory.

There are several further directions which would be interesting
pursue. Firstly, it would be nice to calculate the form factors for
more general final states, i.e. final states involving general numbers
of fermions and scalars. Such expressions should be able to be
compactly expressed in terms of ``super form factors''. At tree-level
this should be accomplished by the application of the BCFW recursion
relations \cite{Gang:2010gy} and use of the superamplitudes
\cite{Gang:2010gy,Bargheer:2010hn} already developed/computed for ABJM.

The duality between colour and kinematics in scattering amplitudes
\cite{Bern:2008qj} has also been discovered to extend to form factors
of ${\cal N}=4$ SYM \cite{Boels:2012ew}. Recently a proposal for such
a duality has been made for the scattering amplitudes of ABJM
\cite{Bargheer:2012gv}. Thus, it would be very interesting to attempt
to extend this to the form factors considered here.

There is also the question of equality between the form factors and
open periodic light-like Wilson loops, developed for the case of
${\cal N}=4$ SYM in \cite{Alday:2007he,Brandhuber:2010ad}. Here there
are subtleties owing to the lack of gauge invariance of the open
Wilson loop, and a gauge-invariant statement is still
lacking\footnote{I thank G. Travaglini for a discussion on this
  point.}. However, it is still worth pursuing this calculation, which
has recently been sharpened in the closed-loop case in
\cite{Bianchi:2013pva}.

Finally, as emphasized in the introduction, the form factors presented
here should be useful for calculating the correlation functions of the
chiral primary operators, following the program of generalized
unitarity as recently proposed in \cite{Engelund:2012re}. The
calculation of the three-point functions at leading order is under
active investigation by the author.

\section*{Acknowledgements}

The author thanks Johannes Henn, Gregory Korchemsky, Charlotte
Kristjansen, Tristan McLoughlin, Jan Plefka, and Gordon Semenoff for
discussions, and also thanks A. Brandhuber, \"O. G\"urdo\v{g}an,
D. Korres, R. Mooney and G. Travaglini for discussions and for sharing
their manuscript prior to publication. The author also thanks
J. A. Minahan, O. Ohlsson Sax, and C. Sieg for permission to use a
figure from their paper. The author was supported in part by FNU
through grant number 272-08-0329.

\appendix
\section{Calculational details}
\label{sec:app}

We follow very closely the conventions of \cite{Minahan:2009wg},
however we evaluate the form factor in Euclidean signature. The regularization
scheme is dimensional regularization with all products involving the
three-dimensional Levi-Civita tensor reduced to scalar products prior
to integration, which is known to be a healthy scheme
\cite{Chen:1992ee}.

The action used in \cite{Minahan:2009wg} is in Lorentzian mostly-positive
signature, with the Levi-Civita tensor defined as $\e^{012} = 1$ and
$\g^\m\g^\n=\eta_{\m\n}+\e_{\m\n\r}\g^\r$
\bsp\label{ABJac}
S = \frac{k}{4\pi}\Tr \int d^3 x\, \Biggl[
&\e^{\m\n\r}\left( A_\m \p_\n A_\r +\frac{2i}{3} A_\m A_\n
A_\r\right)
-\e^{\m\n\r}\left( \hat A_\m \p_\n \hat A_\r +\frac{2i}{3} \hat A_\m
\hat A_\n \hat A_\r\right)\\
&-D_\m Y^{\dag}_A D^\m Y^A + i \psi^{\dag A} \g^\m D_\m \psi_A\\
&+\frac{1}{12}Y^AY_B^\dag Y^CY_D^\dag Y^EY_F^\dag
(\d_A^B\d_C^D\d_E^F
+\d_A^F\d_C^B\d_E^D
-6\d_A^B\d_C^F\d_E^D
+4\d_A^D\d_C^F\d_E^B)\\
&-\frac{i}{2}(
Y_A^\dag Y^B\psi^{\dag C}\psi_D
-\psi_D\psi^{\dag C}Y^BY_A^\dag)
(\d_B^A\d_C^D-2\d_C^A\d_B^D)\\
&+\frac{i}{2}\e^{ABCD}Y_A^\dag\psi_BY_C^\dag\psi_D
-\frac{i}{2}\e_{ABCD}Y^A\psi^{\dag B}Y^C\psi^{\dag D}\Biggr],
\end{split}
\ee
where $D_\m Y^A = \p_\m Y^A + iA_\m Y^A - i Y^A \hat A_\m$ and $D_\m
Y^\dag_A = \p_\m Y^\dag_A - i Y^\dag_A A_\m + i \hat A_\m Y^\dag_A$
and similarly for the fermions. Further details about the action are
available in \cite{Minahan:2009wg}.

The effective Feynman rules are reproduced below, with permission from
\cite{Minahan:2009wg}. 
\begin{center}
\includegraphics[bb=95 438 525 695,clip=true]{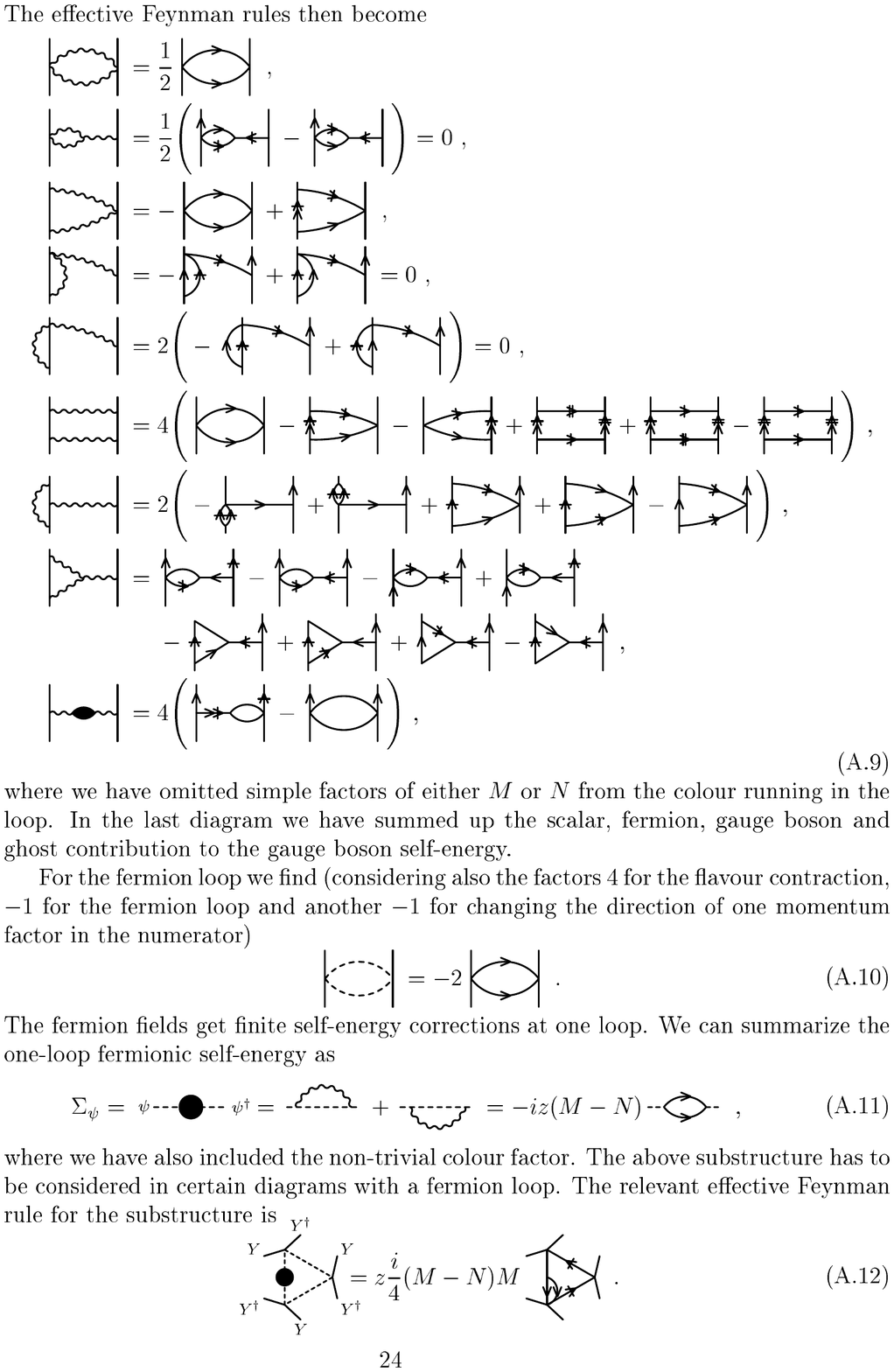}\\
\includegraphics[bb=95 337 525 372,clip=true]{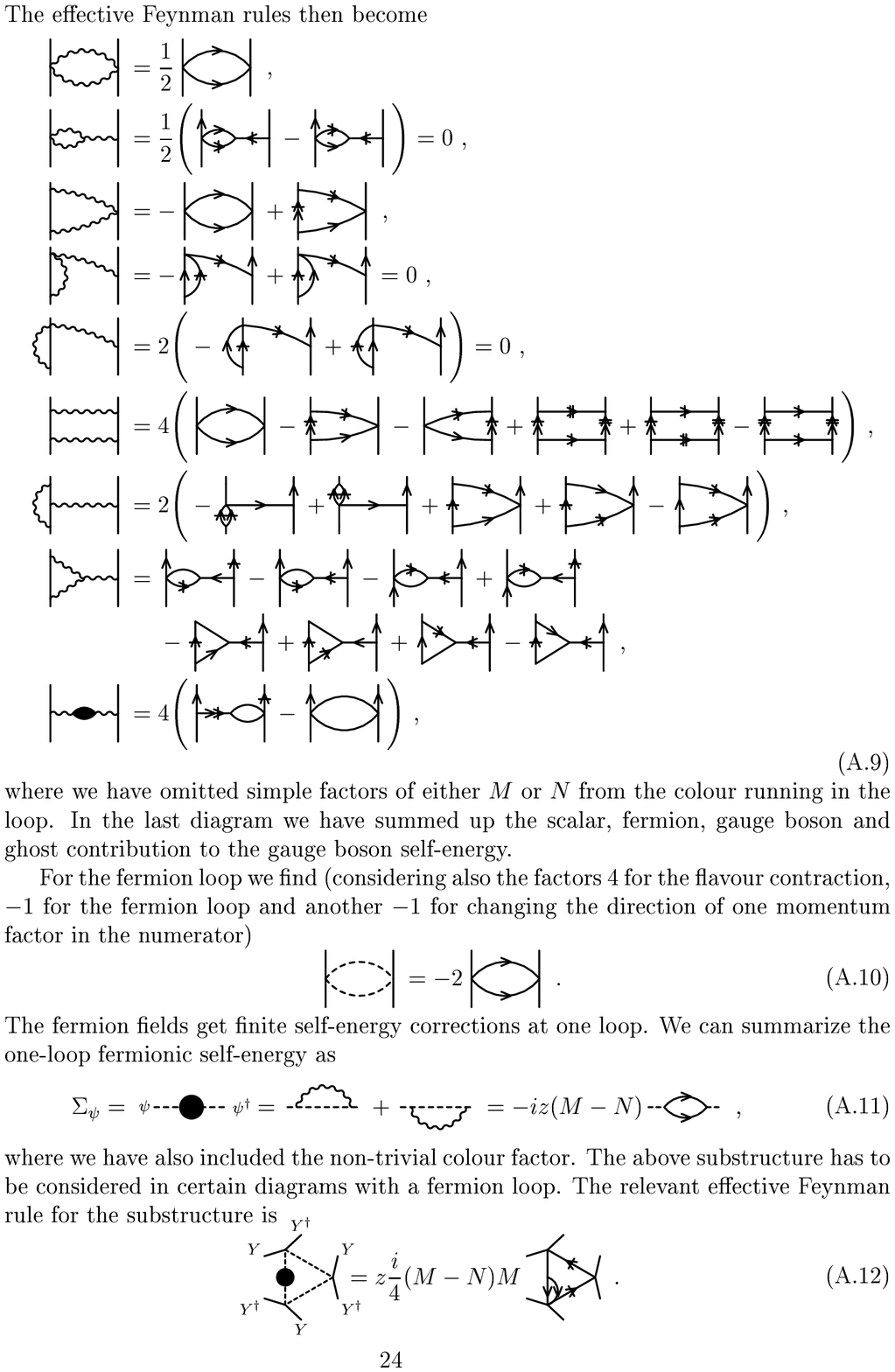}\\
\hspace{-4.05in}\includegraphics[bb=239 202 365 237,clip=true]{mssrules}\\
\includegraphics[bb=95 210 520 290,clip=true]{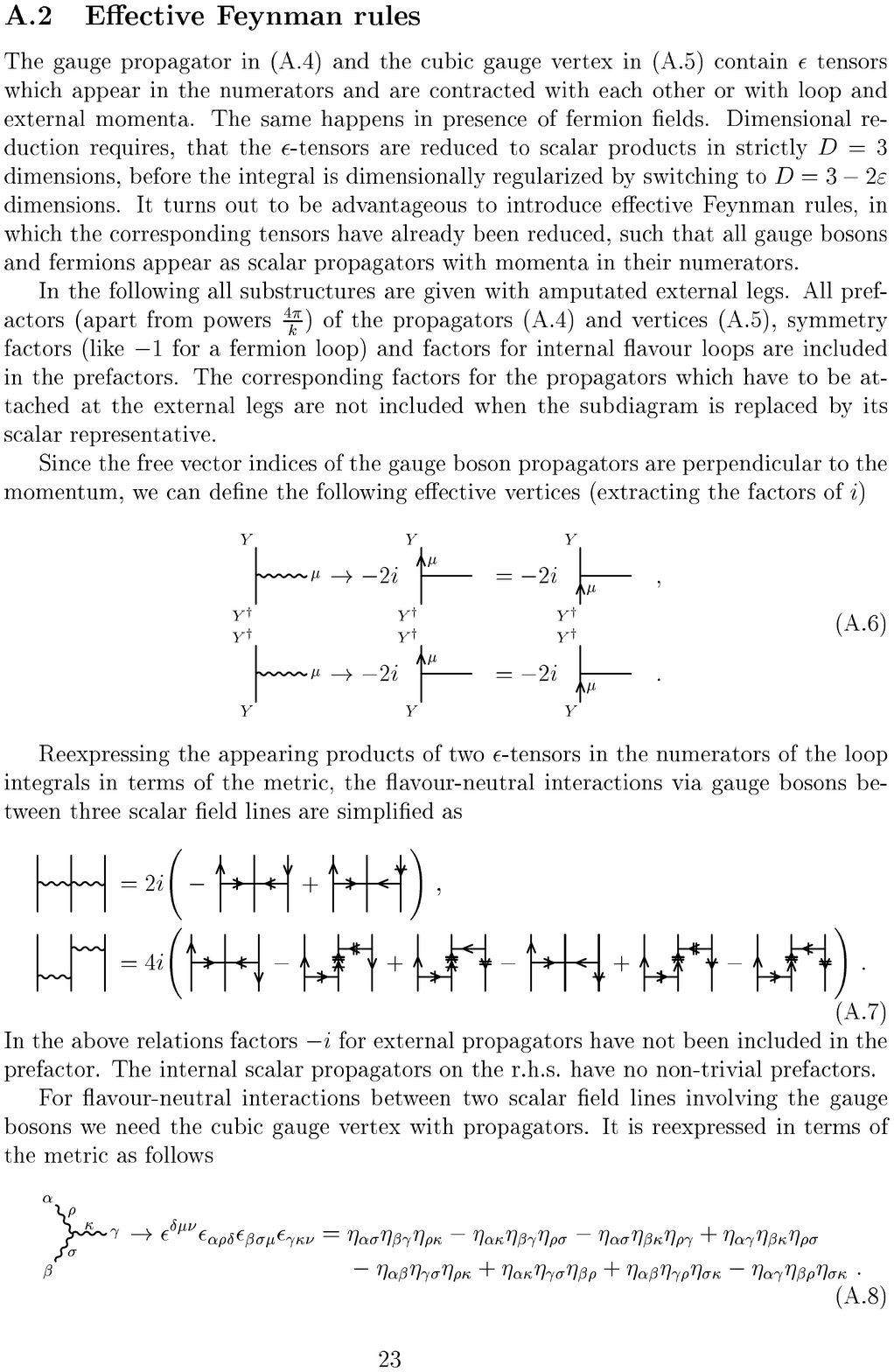}
\end{center}
Gluons are represented by wiggly lines, scalars
by plain lines, and fermions by dashed lines. The filled blob
corresponds to the 1-loop correction to the gluon propagator. On the
right hand side are the effective rules, where arrows appear in pairs
and indicate contracted momenta of the two associated lines. When two
gluons appear on opposite sides of a scalar line, they belong to
opposite gauge groups. 

The procedure for translating these rules to Euclidean signature is to
multiply by $-i$ for every propagator on the left hand side and to
divide by $-i$ for every vertex on the left hand side. This just
removes the $i$'s from the last two rules above and leaves all other
rules unaffected.

The overall coupling assumed in these rules is $4\pi/k$, and hence the
two-loop calculation of the form factors presented here require
multiplication by $(4\pi/k)^2$ times the relevant colour factors.

\subsection{Effective rule for $I_4$}

We find that the effective rule given in \cite{Minahan:2009wg} for
$I_4$ gives inconsistent results in our case. This is likely because
of $0/0$ limits when the bottom two legs are taken on-shell (the
calculation in \cite{Minahan:2009wg} is for an off-shell
quantity)\footnote{I thank C. Sieg for discussions on this
  point.}. For this reason we prefer to use the direct evaluation
given below
\begin{center}
\parbox{20mm}{
\begin{fmfgraph*}(34,80)
\fmftop{v1,v4}
\fmfleft{v2}
\fmfright{v3}
\fmf{fermion,label=$q_1$}{v1,vc1}
\fmf{fermion,label=$q_2$}{vc1,vc4}
\fmf{fermion,label=$q_3$}{vc4,v2}
\fmf{fermion,label=$q_5$}{vc2,v4}
\fmf{fermion,label=$q_4$}{v3,vc2}
\fmffreeze
\fmf{wiggarr,left=0.0,label=$q_6$}{vc1,vc5}
\fmf{wiggarr,right=0.0,label=$q_7$}{vc4,vc5}
\fmf{wiggarr,right=0.0,tension=2,label=$q_8$}{vc2,vc5}
\fmffreeze
\fmfposition
\end{fmfgraph*}}
\end{center}
\vspace{-4cm}
\be\nonumber
=\frac{1}{4} (q_1+q_2)^\m (q_2+q_3)^\n (q_4+q_5)^\l q_6^\phi\,
q_7^\xi\, q_8^\o\,
\e^{\a\b\g}\,\e_{\m\phi\a}\,\e_{\n\xi\b}\,\e_{\l\o\g},
\ee
where the result is understood to be multiplied by the propagators of
the off-shell lines and integrated over the loop momenta. The
contractions of the epsilon tensors must be replaced by scalar
products prior to integration.

\subsection{Effective rule for $I_8$}
\label{sec:appi8}

An effective Feynman rule for the crossed-topology diagram $I_8$ is
not provided in \cite{Minahan:2009wg}. Rather than attempt to derive
one ourselves we decided to simply use the direct evaluation of the
numerators. The result is as follows, where the $q_i$ momenta refer
to the labels in the master topology, which is reproduced from the
main text below
\begin{center}
\vspace{1cm}
\hspace{-3cm}
\parbox{20mm}{
\begin{fmfgraph*}(55,35)
\fmfleft{i1,i2}
\fmfright{o1,o2}
\fmf{plain}{v1,i1}
\fmf{plain,tension=.01,label=$4$,l.side=right,label.dist=-17}{v1,v4c}
\fmf{plain}{v3,o1}
\fmf{plain}{o2,v4}
\fmf{plain,tension=.2,label=$1$,l.side=right}{v4,v2}
\fmf{plain}{i2,v2}
\fmf{plain,tension=.2,label=$2$,l.side=right}{v2,v1c}
\fmf{plain,tension=.2,label=$6$,l.side=right}{v1c,v1}
\fmf{plain,tension=.2,label=$7$,l.side=right}{v3,v4c}
\fmf{plain,tension=.2,label=$3$,l.side=right}{v4c,v4}
\fmf{plain,tension=.01,label=$5$,l.side=left,label.dist=10}{v1c,v3}
\end{fmfgraph*}}
\end{center}\vspace{1cm}
\be\nonumber
I_8 = \frac{1}{4} \Bigl[(q_2+q_6) \w q_4 \w (q_7 -p_2)\Bigr]
 \Bigl[(p_1+q_6) \w q_5 \w (q_7 +q_3)\Bigr] \,G(0,1,1,1,1,1,1),
\ee
where the triple wedge-product indicates contraction with the Levi-Civita
tensor. The product of two such triple products must be re-expressed
as scalar products of momenta prior to evaluation using the usual identity.

\subsection{On the vanishing of certain diagrams}
\label{sec:appv}

There is a simple way to see that the following diagrams are zero on
the physical dimension (and so are at most ${\cal O}(\e)$ in
dimensional regularization; we have verified that they are, in fact,
exactly zero)
\begin{center}
\parbox{20mm}{\vspace{1cm}
\begin{fmfgraph*}(17,40)
\fmftop{v1}
\fmfleft{v2}
\fmfright{v3}
\fmfv{decoration.shape=circle,decoration.filled=30}{v1}
\fmf{plain}{v1,vc1}
\fmf{plain}{vc1,vc3}
\fmf{plain}{vc3,v2}
\fmf{plain}{v1,vc2}
\fmf{plain}{vc2,vc4}
\fmf{plain}{vc4,v3}
\fmfdot{vc1,vc2}
\fmfv{label=$x$,label.angle=180}{vc1}
\fmfv{label=$y$,label.angle=0}{vc2}
\fmffreeze
\fmf{photon,right=0.15}{vc1,vc2}
\fmffreeze
\fmfposition
\end{fmfgraph*}}
\hspace{1cm}
\parbox{20mm}{\vspace{1cm}
\begin{fmfgraph*}(17,40)
\fmftop{v1}
\fmfleft{v2}
\fmfright{v3}
\fmfv{decoration.shape=circle,decoration.filled=30}{v1}
\fmf{plain}{v1,vc1}
\fmf{plain}{vc1,vc3}
\fmf{plain}{vc3,v2}
\fmf{plain}{v1,vc2}
\fmf{plain}{vc2,vc4}
\fmf{plain}{vc4,v3}
\fmfdot{vc1,vc2}
\fmfv{label=$x$,label.angle=180}{vc1}
\fmfv{label=$y$,label.angle=0}{vc2}
\fmffreeze
\fmf{photon,right=0.15}{vc1,vc2}
\fmf{photon,right=0.15}{vc3,vc4}
\fmffreeze
\fmfposition
\end{fmfgraph*}}
\hspace{1cm}
\parbox{20mm}{\vspace{-1cm}
\begin{fmfgraph*}(22,20)
\fmftop{v1}
\fmfstraight
\fmfbottom{v2,v3,v4}
\fmfv{decoration.shape=circle,decoration.filled=30}{v1}
\fmf{plain}{v1,vc1}
\fmf{plain}{vc1,v2}
\fmf{plain}{v1,vc2}
\fmf{plain}{vc2,vcc2}
\fmf{plain}{vcc2,v3}
\fmf{plain}{v1,vc3}
\fmf{plain}{vc3,v4}
\fmffreeze
\fmfdot{vc1,vc2}
\fmfv{label=$x$}{vc1}
\fmfv{label=$y$,label.angle=-40,label.dist=4}{vc2}
\fmf{photon}{vc1,vc2}
\fmf{photon}{vcc2,vc3}
\end{fmfgraph*}}
\vspace{-1cm}
\end{center}
where we can also imagine an arbitrary number of additional legs
emanating from the operator -- the legs shown are assumed to be
adjacent. We consider the one-gluon exchange across two
legs of the operator, and we work in position space. To illustrate we
have marked the two ends of the gluon exchange with their space-time
coordinates in the diagrams above. We take the operator to lie at the
origin. The gluon exchange will contribute the following expression
\be
\text{gluon exchange} \propto \p_x P(x) \w \p_x P(x-y) \w \p_y P(y)
\ee
where the triple-wedge product indicates contraction with the
Levi-Civita tensor, and $P(z)\sim 1/z$ is the
position-space scalar propagator. The partial derivatives come from the
scalar couplings to the gluon and from the gluon propagator. This
expression vanishes since it is proportional to $x\w (x-y)\w y$.

\end{fmffile}
\bibliography{ff}

\end{document}